\documentclass[useAMS,usenatbib]{mnras}

\usepackage{graphicx}
\usepackage{epstopdf}
\usepackage{amsmath,esint}

\title[M82 X-2 on its Propeller Line]{The X-ray Pulsar M82 X-2 on its Propeller Line}

\author[D. M. Christodoulou et al.]{D. M. Christodoulou,$^{1,2}$\thanks{E-mail: dimitris\_christodoulou@uml.edu} 
D. Kazanas,$^{3}$\thanks{E-mail: demos.kazanas@nasa.gov}
and S. G. T. Laycock$^{1,4}$\thanks{E-mail: silas\_laycock@uml.edu}
\\
$^{1}$Lowell Center for Space Science and Technology, University of Massachusetts Lowell, Lowell, MA, 01854, USA\\
$^{2}$Department of Mathematical Sciences, University of Massachusetts Lowell, Lowell, MA, 01854, USA\\
$^{3}$NASA Goddard Space Flight Center, Laboratory for High-Energy Astrophysics, Code 663, Greenbelt, MD 20771, USA\\
$^{4}$Department of Physics \& Applied Physics, University of Massachusetts Lowell, Lowell, MA, 01854, USA
}

\begin{document}

\def\gsim{\mathrel{\raise.5ex\hbox{$>$}\mkern-14mu
                \lower0.6ex\hbox{$\sim$}}}

\def\lsim{\mathrel{\raise.3ex\hbox{$<$}\mkern-14mu
               \lower0.6ex\hbox{$\sim$}}}

\pagerange{\pageref{firstpage}--\pageref{lastpage}} \pubyear{2016}

\maketitle

\label{firstpage}

\begin{abstract}

{\it NuSTAR} has detected pulsations from the ultraluminous X-ray source X-2 in M82 and archival {\it Chandra} observations have given us a good idea of its duty cycle. The newly discovered pulsar exhibited at least 4 super-Eddington outbursts in the past 15 years but, in its lowest-power state, it radiates just below the Eddington limit and its properties appear to be typical of high-mass X-ray binaries. M82 X-2 has been described as a common neutron star with a 1~TG magnetic field that occasionally accretes above the Eddington rate and as a magnetar-like pulsar with a 10-100~TG magnetic field that reaches above the quantum limit. We argue in favor of the former interpretation. Using standard accretion theory and the available observations, we calculate the stellar magnetic field of this pulsar in two independent ways and we show that it cannot exceed 3~TG in either case. We discuss the implications of our results for other ultraluminous X-ray sources that occasionally exhibit similar powerful outbursts.

\end{abstract}




\section{Introduction}\label{intro}

Ultraluminous X-ray (ULX) sources are extragalactic compact accreting objects 
characterized by super-Eddington luminosities ($L_X\sim 10^{39-41}$~erg~s$^{-1}$)  
and unusual soft X-ray spectra with blackbody emission around 
$\lsim$0.3~keV and a
downturn above $\sim$5~keV \citep{gla09,fen11,mot14}.
The extreme luminosities can be understood either
as emission at the Eddington limit 
($L_X\lsim L_{\rm Edd} = 1.3\times 10^{38}~M/M_\odot$~erg~s$^{-1}$) 
from intermediate-mass ($M\sim 10^{2-4} M_\odot$) black holes or 
as anisotropic emission with $L_X > L_{\rm Edd}$ from stellar-mass
black holes and neutron stars \citep{sor07,med13,bac14,mot14,pas14}.

The latter interpretation is gaining favor from recent observations: 
\begin{itemize}
\item[(a)]\cite{lua14} analyzed a {\it Chandra} sample of nearby
ULX sources and found a change in the spectral index around 
$L_X\sim 2\times 10^{39}$~erg~s$^{-1}$ that may indicate 
a transition to the super-Eddington accretion regime by 10$M_\odot$ black holes.
\item[(b)]\cite{mot14} determined that the ULX source P13 in NGC7793 harbors
a stellar-mass black hole with $M < 15~M_\odot$.
\item[(c)]\cite{bac14} 
determined that the ULX source X-2 in M82 harbors a pulsar with spin period 
$P_S = 1.3725$~s and spinup rate $\dot{P_S}= -2\times 10^{-10}$~s~s$^{-1}$.
\end{itemize}
In all of these cases, the observations indicate that the X-ray sources radiate above 
their Eddington limits, a result that makes uncomfortable many researchers.

The magnetic field of NGC7793 P13 has not been estimated, but in the case
of M82 X-2, \cite{bac14} used their measurement of the accretion torque
to obtain a modest value of the magnetic field
$B\approx 1$~TG. In contrast, \cite{eks15} argued that the accretion
torque is very small near spin equilibrium and that leads to an underestimate
of the magnetic field which in M82 X-2 must be of magnetar strength ($B\sim 100$~TG)
\citep[see also][]{ho14,klus14,lyu14,dal15,tsy16,dal16}.
We, on the other hand, find that the magnetic field implied from the {\it Chandra}
measurements of M82 X-2 \citep{bri16,tsy16}
is modest ($B\approx 3$~TG), a result that is in agreement with the independent
determination of \cite{bri16} 
based on {\it NuSTAR} observations
(a high-energy cutoff at $14_{-3}^{+5}$~keV).
No unusual physics is necessary to obtain this estimate, the result is derived in two different ways 
from standard accretion theory and the measured values of $P_S$, $\dot{P_S}$, and minimum $L_X$ in the lowest-power states. 
More importantly, such a modest $B$-value is
consistent with isotropic X-ray emission near the Eddington limit 
when the pulsar lies in its low-power accreting state 
\citep[on the so-called ``propeller line;''][]{chr16}.
This consistency implies that M82 X-2 is a common pulsar whose
propeller line lies near the Eddington limit 
($L_X\approx L_{\rm Edd}$) and
that occasionally bursts anisotropically in our direction, an explanation also favored by \cite{bac14}. On the other hand, \cite{tsy16} and \cite{dal16} have interpreted the low-$L_X$ states of M82 X-2 
as quiescent magnetospheric emission that occurs when the system
straddles the \cite{cor96} gap. We show below that this interpretation is not supported by the observations.

The above results are already present in the equations
of accretion on to compact objects \citep{els77,gho79,wan87,wan95,wan96,fra02}, but the physics is obscured
by the multiple scalings and some minor inconsistencies in the definitions of the physical
quantities involved. We thus derive and discuss the accreting pulsar's minimum isotropic
luminosity and surface magnetic field in \S~\ref{eqs} without 
scaling the various physical variables, and we discuss our conclusions for M82 X-2
and other ULX sources in \S~\ref{conclusions} below.

\section{Equations for Accretion on to a Compact Object}\label{eqs}

We adopt Gaussian units for electromagnetic quantities \citep{jac62}
and we consider accretion from a disk-like structure that is formed 
around a compact object. As in \cite{pri72}, we assume that
on the propeller line
the inner edge of the disk reaches inward to corotation 
(eqs.~[\ref{td1}] and~[\ref{rc}] below) and this
condition is sufficient to determine the lowest emitted X-ray luminosity due to accretion. But we also
consider the location of the magnetospheric radius 
(eqs.~[\ref{td2}] and~[\ref{rm2}] below) in order to additionally
derive an estimate of the stellar magnetic field as was done by
\cite{gho79} and \cite{fra02}. 

We introduce below three control parameters in order to bridge the differences between many calculations carried out in the past: (a)~parameter
$\xi$ specifies the location of the disk's inner edge relative to the location of the spherical magnetospheric radius \citep{gho79,wan96}; (b)~parameter $\alpha$ unifies two different definitions of the magnetic moment $\mu$
\citep[e.g.,][]{wan96,eks15}; and (c)~$\eta$ is the efficiency of converting accretion power to X-ray luminosity.

\subsection{Accretion Torque}\label{tau1}

The accretion torque at the inner edge of the disk can be written as
\begin{eqnarray}
 \tau_{d} &=&  \dot{M} \sqrt{G M R_{d}}  \label{td1} \\
               &=&  \dot{M} \sqrt{G M (\xi r_m)} \label{td2} \, ,
\end{eqnarray}
where $\dot{M}$ is the mass accretion rate at the inner edge of the disk $R_{d}$, 
$G$ is the gravitational constant, $M$ is the mass of the compact object,  
and $r_m$ is the spherical magnetospheric (Alfv\'en) radius. 
We assume that the accreted matter 
is forced to follow magnetic field lines at the cylindrical radius $R_d$
and we adopt the relation 
$R_{d} \equiv \xi r_m$ \citep{wan96}.

The accretion torque can be measured 
\citep{bac14} as it can also be be expressed in terms of the transfer of 
angular momentum onto the compact object during spinup:
\begin{equation}
|\tau_*| = |\dot{\cal L}_*| = \frac{2\pi I_*}{P_S^2} |\dot{P_S}| \, ,
\label{tau}
\end{equation}
where ${\cal L}_* = 2\pi I_*/P_S$ and $I_*$ are the angular momentum
and the moment of inertia, respectively, of the compact object with spin period $P_S$ and derivative $\dot{P_S}$. Once a nonzero $\tau_*$ has been measured, the torque in eq.~(\ref{td1}) can no longer be adjusted: any attempt to lower $\tau_d$ using the ``fastness parameter'' \citep{eks15,dal15} leads to a corresponding increase in $\dot{M}$ that
maintains conservation of angular momentum. This pushes $L_X$ higher and the increased luminosity no longer describes the propeller line.

\subsection{Magnetospheric Radius}\label{ra}

Calculations involving the spherical magnetospheric (Alfv\'en) radius $r_m$ 
in the literature \citep{els77,gho79,wan96,fra02,eks15}
suffer from inconsistencies in the definitions of the 
magnetic moment $\mu$ and the ram pressure ${\cal P}_{ram}$ of the inflowing matter.
Eq.~(11) of \cite{gho79} agrees with eq.~(6) of \cite{els77}
only if ${\cal P}_{ram}=\rho v^2 / 2$ in the former calculation, where $\rho$ is the mass 
density and $v$ is the inflow speed of matter at spherical radius $r$.
In both cases, the magnetic moment is defined as $\mu = B r^3$ on dimensional grounds \citep[see also][]{wan96}, 
where $B$ is the magnetic field at $r$;
but other authors use $\mu = B r^3 / 2$ instead \citep[e.g.,][]{eks15}.
The factor of 1/2 alters the definition of $r_m$ but the modification
has not been implemented by other authors who continue to use the classical formula for $r_m(\mu)$.
Thus, various factors of 1/2 propagate in the calculations but they cause small differences in the coefficients of the final results because of the steep 
dependence of $B^2(r)$ on $1/r^6$. For example, using the 1/2 in the
definition of $\mu$ would expand $r_m(\mu)$ by a factor of $16^{1/7}=1.486$. We account for this discrepancy below by introducing the control parameter $\alpha$ in the definition of $\mu$.

We adopt the following definitions:~for the ram pressure, 
${\cal P}_{ram}=\rho v^2 / 2$ 
\citep[consistent with Bernoulli's equation;][]{wan96,men99};
and for the magnetic moment, $\mu = \alpha B r^3$, where $\alpha$
can be taken as 1 or 1/2.
Then $\mu_*/\alpha = B r^3 = B_* R_*^3$, where asterisks indicate the quantities at the
surface of the compact object, and the magnetic pressure at radius $r$ can be written as 
${\cal P}_{mag} = B^2/(8\pi) = \mu_*^2/(8\pi\alpha^2 r^6)$. 
As usual 
\citep[e.g.,][]{fra02},  
$\rho |v| = \dot{M}/(4\pi r^2)$ and $|v|$ is close to the free-fall speed
$v_{\rm ff} = \sqrt{2GM/r}$, in which case the balance between the two pressures 
at $r=r_m$ determines the spherical Alfv\'en radius:
\begin{equation}
r_m = \left(\frac{1}{\alpha^4}\frac{\mu_*^4}{2 G M \dot{M}^2}\right)^{1/7} \, .
\label{rm}
\end{equation}
The classical formula is recovered for $\alpha = 1$.
It is convenient to replace $\dot{M}$ in terms of the X-ray luminosity $L_X$,
which is assumed to be a fraction $\eta$ of the total accretion power, i.e., 
\begin{equation}
L_X = \eta\left(\frac{G M \dot{M}}{R_*}\right) \, ,
\label{lacc}
\end{equation}
in which case eq.~(\ref{rm})
takes the form
\begin{equation}
r_m = \left(\frac{\eta^2}{2\alpha^4}
\frac{G M \mu_*^4}{L_X^2 R_*^2}\right)^{1/7} \, .
\label{rm2}
\end{equation}

\subsection{Minimum Isotropic X-Ray Luminosity}\label{lm}

Assuming that $|\tau_*| = \tau_{d}$, we get from eqs~(\ref{td1}), (\ref{tau}), and~(\ref{lacc}):
\begin{equation}
L_X = \eta\left(\frac{2\pi I_*}{R_* P_S^2} |\dot{P_S}|\right)\sqrt{\frac{G M}{R_{d}}} \, .
\label{lmin1}
\end{equation}
If we ask for the inflowing matter that reaches $R_{d}$ to also corotate
with the magnetic field lines, then the inner-disk radius $R_{d}$ must be equal to the corotation 
radius $r_c$:
\begin{equation}
R_{d} = r_c \equiv \left(\frac{G M P_S^2}{4\pi^2}\right)^{1/3} \, .
\label{rc}
\end{equation}
Eliminating $R_{d}$ between eqs.~(\ref{lmin1}) and~(\ref{rc}), we find that the minimum X-ray luminosity on the propeller line is
\begin{equation}
L_X = \eta\left(2\pi I_* |\dot{P_S}|\right)\left( \frac{2\pi}{P_S^7}\frac{G M}{R_*^3} \right)^{1/3} \, .
\label{lmin2}
\end{equation}
The same equation (without the $\eta$) was also derived by \cite{gal08} who used it to obtain estimates of the minimum X-ray luminosity in a large sample of Magellanic high-mass X-ray binaries with measured $P_S$ and $\dot{P_S}$ values.

It is convenient to rewrite the result in terms of the angular velocity of the
compact object $\Omega_S\equiv 2\pi /P_S$ and the Keplerian angular velocity
on the surface $\Omega_{K*}\equiv\sqrt{G M / R_*^3}$~:
\begin{equation}
L_X = \eta ~I_* \left( \Omega_S \Omega_{K*}^2 \right)^{1/3} |\dot{\Omega}_S| \, .
\label{lmin3}
\end{equation}
The cubic root in this equation is the geometric mean of $\Omega_S$ 
and $\Omega_{K*}^2$. Thus, the luminosity $L_{X}\propto |\dot{\Omega}_S|$
and the proportionality constant is weighted twice by $\Omega_{K*}$, the larger
of the two frequencies in the geometric mean, which also expresses the depth
of the gravitational potential well of the compact object. Despite that,
it is well-known that
the geometric mean favors the smaller values, therefore it is the spin
frequency $\Omega_S$ that dominates the proportionality constant.
This implies that a source that is
spun up by accretion will increase its power output only marginally  
($L_{X}\propto \Omega_S^{1/3}$) and corotation will move inward
very slowly.
For a compact object spinning near breakup ($\Omega_S = \Omega_{K*}$), 
corotation is near the surface of the object ($r_c \approx R_*$) and eq.~(\ref{lmin3})
reduces to $L_{X} = \eta I_* \Omega_{K*} |\dot{\Omega}_K|$ which is simply 
the rate of radiative loss of a fraction $\eta$ of the rotational kinetic energy 
$E_{rot}=I_* \Omega_{K*}^2 /2$ that reaches the surface.

\subsection{Stellar Magnetic Field}\label{bm}

We set again $|\tau_*| = \tau_{d}$ and we combine eqs~(\ref{td2}), (\ref{tau}), 
(\ref{lacc}), (\ref{rm2}), and (\ref{lmin2}) 
to get a relation that involves  the stellar magnetic moment.
We find that
\begin{equation}
\mu_{*} = \alpha\left(2\pi^2 \xi^7\right)^{-1/4} \sqrt{G M I_* |\dot{P_S}|} \, ,
\label{mu1}
\end{equation}
or, using $\mu_{*} = \alpha B_{*} R_*^3$, that the surface magnetic field is independent of $\alpha$:
\begin{equation}
B_{*} = \left(2\pi^2 \xi^7\right)^{-1/4}
\sqrt{\frac{G M I_*}{R_*^6} |\dot{P_S}|} \, .
\label{b1}
\end{equation}
The results in eqs.~(\ref{mu1}) and~(\ref{b1}) do not depend on the precise amount 
of the inflowing accretion power that is converted to X-rays. Therefore, the efficiency
of the conversion process (expressed by $\eta$ in eq.~[\ref{lacc}]) does not enter
in the calculation of $B_*$. This result makes physical sense.
Eq.~(\ref{b1}) also makes sense: it relates the stellar magnetic field $B_*$ to nonmagnetic physical variables (see below), therefore
$B_*$ is independent of $\alpha$.

Written in terms of $\Omega_S$ and $\Omega_{K*}$, eq.~(\ref{b1}) becomes
\begin{equation}
B_{*} = \left( \frac{\xi^7}{2} \right)^{-1/4} \left(\frac{\Omega_{K*}}{\Omega_S}\right) 
\sqrt{\frac{I_*}{R_*^3} |\dot{\Omega}_S|} \, .
\label{b2}
\end{equation}
This equation relates the surface magnetic pressure to
the $R\phi$ component of the stress tensor $|\rho v_R v_K|_{_{R_{d}}}$ at $R_{d}$,
where $\rho (R_{d}) |v_R(R_{d})| = \dot{M}/(4\pi R_{d}^2)$ and
$v_K(R_{d}) = (GM/R_{d})^{1/2}$~: squaring both sides of eq.~(\ref{b2}), 
we find that
\begin{equation}
\frac{B_{*}^2}{8\pi} = \left( 2 \xi^7 \right)^{-1/2} 
\left(\frac{\Omega_{K*}}{\Omega_S}\right)^2 \left(\frac{R_{d}}{R_*}\right)^3 
|\rho v_R v_K|_{_{R_{d}}}\, .
\label{b3}
\end{equation}
This is because the torque balance
$|\tau_*|= \tau_{d}$ implies that
\begin{equation}
\frac{I_*}{R_*^3} |\dot{\Omega}_S| = 4\pi ~|\rho v_R v_K|_{_{R_{d}}} 
\left(\frac{R_{d}}{R_*}\right)^3 \, .
\label{ior}
\end{equation}
Eq.~(\ref{b3}) shows that the stress tensor at $R_{d}$ is strongly rescaled 
by both the spin frequency of the compact object (ratio ($\Omega_{K*} / \Omega_S)^2 >> 1$) 
and by the scale of the accretion flow (ratio ($R_{d} / R_*)^3 >> 1$) in order to match the 
magnetic pressure on the surface of the compact object.
Furthermore, eq.~(\ref{b3}) reveals the following two notable accretion properties:
(a)~If the corotation radius $R_{d}$ is smaller, then the magnetic field 
is weaker and vice versa, precisely as expected. (b)~If the compact object is spun up
($\Omega_S$ increases), then the magnetic field appears to be weaker and vice versa, 
again as expected.
The two effects do not work against one another because the direction in which the two ratios 
move during spinup or spindown is the same. 

Let us now imagine that the compact object is spun up from accretion at a certain moment
in time. Then $P_S$ decreases, $\Omega_S$ increases, and corotation pushes inward 
(eq.~[\ref{rc}]). This results in a decrease of both ratios in eq.~(\ref{b3}) and then, 
since $B_{*}$ remains unchanged on the surface, the $R\phi$ component of the stress 
tensor at $R_{d}$ in eq.~(\ref{b3}) must necessarily increase. 
Since the condition $|\tau_*| = \tau_{d}$ and eqs.~(\ref{td1}) and~(\ref{tau})
imply that $\dot{M}\propto 1/(P_S^2 R_{d}^{1/2})$, then $\dot{M}$ also
increases and this leads to an increase in the X-ray luminosity of the propeller line.

\subsection{Another Form of the X-ray Luminosity on the Propeller Line}\label{pline}

Eliminating $I_* |\dot{P_S}|$ between eqs.~(\ref{lmin2}) and~(\ref{mu1}), we obtain the minimum luminosity on the propeller line \citep[where $r_c=\xi r_m$ as in eq.~(5) of][]{ste86}, that is 
\begin{equation}
L_X = \left(\frac{\eta}{\alpha^2}\right)\left(\frac{\xi^7}{2}\right)^{1/2}
\left(\frac{\Omega_S}{\Omega_{K*}}\right)^{4/3}
\frac{\mu_*^2 \Omega_S}{R_*^3} \ .
\label{pline1}
\end{equation}
The result of \cite{ste86} is reproduced for $\eta = 1 = \alpha$, $\xi=0.5$ (which is reasonable for $L_X << L_{\rm Edd}$), and canonical pulsar parameters. For these choices, 
the leading coefficient in eq.~(\ref{pline1}) is 
$1.4\times 10^{37}~{\rm erg~s}^{-1}$, only by a factor of $1/\sqrt{2}$ lower than
$2\times 10^{37}~{\rm erg~s}^{-1}$ in \cite{ste86}.

The propeller line, eq.~(\ref{pline1}) with $\xi=0.5$, can be written in the physical form
\begin{equation}
L_X = \left(\frac{\eta}{16}\right)B_*^2 R_*^2 v_*
\left(\frac{\Omega_S}{\Omega_{K*}}\right)^{4/3} \ ,
\label{pline2}
\end{equation}
where $v_* \equiv \Omega_S R_* = 2\pi R_*/P_S$. 
The interpretation of this equation is straightforward: 
The isotropic power output that emerges from the surface of the neutron star is scaled down strongly by the ratio $(\Omega_S/\Omega_{K*})^{4/3} << 1$ 
when the source resides on the propeller line and accretion is minimal.

The value of $\xi = 0.5$ used so far is justified by numerical simulations 
of star-disk models in low-power states near the propeller line \citep{rom02,rom03,rom04,lon05,bes08,zan13}. These simulations were designed to investigate protostellar systems and low-luminosity accreting neutron stars and they do not address the question of propeller emission at the Eddington limit ($L_X\approx L_{\rm Edd}$). For such (ULX-type) systems in which minimal accretion proceeds at the Eddington rate, a value of $\xi = 1$ appears to be appropriate. This value is supported by a variety of theoretical calculations \citep{wan87,wan95,wan96,aro93,ost95}.
The stellar magnetic field on the propeller line of such systems is also expected to be stronger than that found on the lowest propeller line of Magellanic pulsars \citep{chr16}, and the stronger magnetic field should not allow the disk to cross inside the spherical magnetospheric radius. For $\xi = 1$ and $\mu_* = \alpha B_* R_*^3$, eq.~(\ref{pline1}) takes a form independent of $\alpha$, that is
\begin{equation}
L_X = \left(\frac{\eta}{\sqrt{2}}\right) B_*^2 R_*^3 \Omega_S \left(\frac{\Omega_S}{\Omega_{K*}}\right)^{4/3} \ ,
\label{pline3}
\end{equation}
and this equation allows for an independent determination of the stellar magnetic field $B_*$ of ULX-type sources from measurements of $\Omega_S$ and the minimum $L_X$ {\it without assuming torque balance}. Only one such ULX source is currently known to harbor a pulsar, M82 X-2, and we provide our estimates for this system in the next subsection.

\subsection{Application to M82 X-2}\label{x2}

The following measurements have been collected from observations of M82 X-2:
\begin{itemize}
\item[(a)]\cite{bac14} measured $P_S=1.3725$~s and $\dot{P_S}=-2\times 10^{-10}$~s~s$^{-1}$ from {\it NuSTAR} observations when the system was in outburst.
\item[(b)]\cite{bri16} analyzed 15 years of {\it Chandra} observations and their results show that the source was sitting in its lowest-power state with $L_{X} = 1.0\times 10^{38}~{\rm erg~s}^{-1}$ in at least two epochs (in 1999 and 2009), and perhaps in February 2013 too.
\item[(c)]\cite{tsy16} analyzed the same {\it Chandra} observations and their results are similar reporting a lowest-power state with $L_{X} = (1.7-2.3)\times 10^{38}~{\rm erg~s}^{-1}$.
\item[(d)]\cite{dal16} found an upper limit of $L_{X} = 1.7\times 10^{38}~{\rm erg~s}^{-1}$ for no detection of the pulsar in a {\it Chandra} observation.
\end{itemize}

We assume that the lowest-power state found in the {\it Chandra} observations occurs when the source lies on its propeller line and we proceed to calculate its minimum X-ray luminosity and the stellar magnetic field. 
We use $G=6.67\times 10^{-8}$~cm$^3$~g$^{-1}$~s$^{-2}$, $\xi = 1$, $I_* = 2 M R_*^2/5$, 
and the following canonical pulsar parameters: 
$M=1.4 M_\odot$ and $R_* = 10$~km.

Using the measured values of $P_S$ and $\dot{P_S}$,
we find from eq.~(\ref{lmin2}) that
\begin{equation}
L_{X} = 7.1\times 10^{38} ~\eta ~~{\rm erg~s}^{-1}\, .
\label{lmins}
\end{equation}
The results of \cite{kor06} suggest that $\eta = 1/4$ is typical for neutron stars. Then, the power output of M82 X-2 on its propeller line appears to be equal to the Eddington limit.

Using the measured value of $\dot{P_S}$ (i.e., torque balance),
we also find from eq.~(\ref{b1}) that
\begin{equation}
B_{*} = 3.1\times 10^{12} ~~{\rm G}\, .
\label{bs}
\end{equation}
But the magnetic field can also be determined independently from the minimum luminosity on the propeller line and without using the value of $\dot{P_S}$. In this case, eq.~(\ref{pline3}) can be written in the form
\begin{equation}
B_{*} = 8.0\times 10^{11} \sqrt{\frac{L_X/(10^{38}~{\rm erg~s}^{-1})}{\eta}}\left(\frac{P_S}{1~{\rm s}}\right)^{7/6} ~~{\rm G}\, .
\label{pline4}
\end{equation}
For $P_S = 1.3725$~s, $L_{X} = 1.0\times 10^{38}~{\rm erg~s}^{-1}$ \citep[based on the results of][]{bri16}, and $\eta = 1/4$ this method yields
\begin{equation}
B_{*} = 2.3\times 10^{12} ~~{\rm G}\, .
\label{bs2}
\end{equation}
Alternatively, for $L_X = L_{\rm Edd} = 1.8\times 10^{38}$~erg~s$^{-1}$ \citep[based on the results of][]{tsy16},
the stellar magnetic field is given again by eq.~(\ref{bs}).
The agreement between the above determinations of $B_*$
justifies the implementation of torque balance in the derivation of eq.~(\ref{bs}).

\cite{tsy16} found a bimodal distribution of X-ray luminosities in the {\it Chandra} data for M82 X-2 with a peak value of
$L_{X} = 1.0\times 10^{40}~{\rm erg~s}^{-1}$ in the high-power state and
$L_{X} = 2.8\times 10^{38}~{\rm erg~s}^{-1}$ in the low-power state. They assumed that the source straddles the \cite{cor96} gap
between the two states and exhibits magnetospheric emission in the low-power state. We do not believe that this assessment of the two states of M82 X-2 is correct. For $P_S=1.3725$~s, the \cite{cor96} gap is given by
\begin{equation}
\Gamma = 167.87~P_S^{2/3} = 207 \, ,
\label{cgap}
\end{equation}
and the low state below the gap is located at a luminosity of
$L_X/\Gamma = 4.8\times 10^{37}~{\rm erg~s}^{-1}$,
if the propeller line is at the high observed state ($L_{X} = 1\times 10^{40}~{\rm erg~s}^{-1}$). Then the low observed state lands well within the gap. Conversely, if the low observed state ($L_{X} = 2.8\times 10^{38}~{\rm erg~s}^{-1}$) is assumed to be at the low boundary of the gap, then the propeller line is located $207$ times higher at $L_{X} = 5.8\times 10^{40}~{\rm erg~s}^{-1}$, i.e., several times higher than the most luminous observations to date \citep{bri16,tsy16}. In this case, the outbursts of the source
land well within the gap. For these reasons, our assumption that the low observed state corresponds to the propeller line and the high observed state corresponds to type-II outbursts \citep{coe10} appears to be more reasonable.

\section{Discussion and Conclusions}\label{conclusions}

We have used standard accretion theory to estimate the typical values of
the minimum isotropic luminosity and the surface magnetic field for the ULX pulsar M82 X-2 for which
\cite{bac14} have recently measured pulsations with period $P_S = 1.3725$~s
and spinup rate $\dot{P_S}=-2\times 10^{-10}$~s~s$^{-1}$.
For these values of $P_S$ and $\dot{P_S}$, our results show that the minimum
luminosity due to accretion (the propeller line) is equal to the Eddington luminosity. We suspect that this also will turn out to be a characteristic property of other ULX sources that harbor pulsars.

X-ray emission at the Eddington level has been previously observed from A0538-66 in the Large Magellanic Cloud \citep{ski82}
and the spin period of the pulsar (69.2~ms) was
then measured. All other X-ray observations of A0538-66 since then, summarized
by \cite{kre04}, have measured no pulsations and much smaller luminosities that fall well 
below the propeller line \citep{ste86,chr16}.
This led \cite{cam95} and \cite{cam97} 
to propose that the low-power X-rays are emitted from the magnetosphere 
where accretion is halted by the centrifugal force and a luminous shock forms.
M82 X-2, for which \cite{kon07} and \cite{dal16} have reported three cases of no detection, 
may also be caught in such nonpulsating sub-Eddington states 
if it will be observed below its propeller line 
($L_X = 1\times 10^{38}~{\rm erg~s}^{-1}$),
and spectral differences (e.g., softer spectra) ought to provide 
an indication that the propeller line has been crossed.
In fact, another ULX source hosting a stellar-mass black hole (NGC7793 P13)
has been observed in such a ``faint" state 
\citep[$L_X = 5\times 10^{37}~{\rm erg~s}^{-1}$;][]{mot14}.

The surface magnetic field of M82 X-2 is estimated to be 
$B_{*}\approx 3~{\rm TG}$ by two independent methods.
One estimate comes from the lowest observed states 
\citep{bri16,tsy16} that effectively define the propeller line.
This method does not use the measured value of the spinup rate $\dot{P_S}$, but it depends on the efficiency $\eta$ of converting accretion power to X-rays (eq.~[\ref{pline4}]).
The second estimate uses torque balance and $|\dot{P_S}|$,
but it does not depend on $\eta$ (eq.~[\ref{bs}]).
The results converge to effectively the same moderate value for $B_*$ despite the uncertainties in the control parameters ($\xi$ and $\eta$) and the direct measurements ($P_S$ and $\dot{P_S}$) involved.

From the above estimates,
we conclude that there is no need to invoke powerful, magnetar-type magnetic
fields to explain the pulsar's X-ray power output on the propeller line, standard accretion
mechanisms suffice. \cite{eks15} and \cite{lyu14} 
advocated for such very strong magnetic
fields ($B\sim 10-100~{\rm TG}$) because they thought that M82 X-2 is 
in spin equilibrium characterized
by a substantial reduction in accretion torque. Our results in \S\S~\ref{lm}$-$\ref{x2}
and the measured spinup rate $\dot{P}_S$ argue against
this claim and against using the peak X-ray luminosity in order to estimate 
the stellar magnetic field. This is also a problem in the calculation of \cite{ton15},
although $L_X$ was arbitrarily reduced from its peak value by an anisotropic factor of
5, causing a reduction to the dipolar $B_*$ by $5^3$ and making it agree with our determination
\citep[see also][where the dipolar field undergoes decay during supercritical accretion and ends up at about the same value]{pan16}.

The studies by \cite{klu15} and \cite{dal15,dal16} 
have assumed that the enormous power output of M82 X-1 is isotropic.
We do not support these scenarios because, if that were the case, the
radiation pressure could easily disrupt the accretion process and the inner disk itself.
This point is made clear by the calculations of \cite{pri72} who found
limits to the mass inflow rate in order for quasistatic accretion to continue uninterrupted
by the radiation pressure.
In the same context, we also do not agree with the argument of \cite{eks15}
and \cite{dal15} that a very strong magnetic field would suppress the
Thomson and Compton scattering cross-sections near the star's surface permitting thus 
accretion to release energy above the conventional Eddington limit. 
We disagree because the magnetic
field decreases quite rapidly with distance from the star and the problem
of super-Eddington X-ray emission would then become acute at a few tens of stellar
radii above the star's surface where the magnetic field becomes weak and the
scattering cross-sections assume again their standard values.

Thus it appears that a
rather large (but theoretically attainable) amount of anisotropy appears to be necessary in order to explain the super-Eddington luminosities
($L_{\rm max}\approx 2\times 10^{40}~{\rm erg~s}^{-1}$)
observed during outbursts \citep{bri16,tsy16}. In these episodic events, the supply of matter (the $\dot{M}$) at corotation increases by a factor of $\sim$2 \citep{bac14} and accretion becomes supercritical. The radiation cannot escape in all directions because the surrounding inflowing stream is optically thick. The photons can only escape in the axial direction where radiation pressure pushes an outflow to speeds larger than the local escape speed. 
The outflow is highly anisotropic and we need to be in a favored position in order to detect the apparent super-Eddington flux liberated in the funnel.
If we assume that a luminosity of $L_{\rm Edd}$ appears to the observer as $L_{\rm max}$, then we find that the opening angle of the emission funnel pointing in our direction is about $6.5^o$. 
The fact that we need to be in such a favored position explains the scarcity of ULX sources.

Furthermore, if the emission from M82 X-2 is collimated, some of the energy must emerge
at much longer wavelengths. This is the case according to the radio observations
of \cite{kro85}, \cite{mcd02}, and \cite{fen08};
and the infrared observations of \cite{kon07} and \cite{gan11}. 
The radio maps show a core-dominated source which is expected if the pulsar is a 
modestly aligned rotator and a collimated jet is coming out in our direction. 
\cite{kon07} also produced {\it Chandra} X-ray spectra that are
hard (photon indices $1.3-1.7$ from an absorbed power-law model) and show no soft excess.
This, combined with the strong X-ray variability on timescales of $\sim$2 months and the 
reccuring type-II outbursts, indicates that M82 X-2 is not at all dissimilar to Galactic and
Magellanic X-ray binaries harboring neutron stars.
The recent observations reported by \cite{bri16} and \cite{tsy16}
have effectively confirmed this picture.

There are some more discoveries about ULX sources indicating that they 
may be supercritical accretors and type-II bursters, but otherwise not at all exotic:  
\begin{itemize}
\item[(a)]\cite{gil04} found that the ULX sources are just the high
end of a luminosity function that cuts off at 
$L_X\sim 3\times 10^{40}~{\rm erg~s}^{-1}$
and in which the known high-mass X-ray binaries make up
the low end of a single power-law with slope $\sim$1.6.
\item[(b)]\cite{lua14} found that the spectral index of ULX spectra changes
around $L_X\sim 2\times 10^{39}~{\rm erg~s}^{-1}$, a value that may indicate
a transition from critical to supercritical accretion by common 10$M_\odot$ black holes.
\item[(c)]\cite{lay15} determined from the radial velocity curve of IC10 X-1
that the the compact object is likely a neutron star, although a low-stellar-mass 
black hole cannot be ruled out.
\item[(d)]\cite{liu13} determined from optical spectroscopy that the mass of the
compact object in M101 ULX-1 is no more than 30$M_\odot$. As in the case of NGC7793 P13 \citep{mot14},
it is unlikely that this is an intermediate-mass black hole.
For this system, \cite{she16} 
developed a model that explains how reprocessed soft X-rays can be finally emitted
from very large radii ($\sim$100 times beyond the inner radius of the accretion disk).
This model may also be applicable to other ULX sources harboring 
stellar compact objects \citep[see also][]{kin08}.
\end{itemize}

Soon after the discovery of pulsations in M82 X-2, \cite{dor15} 
tried to find more pulsating ULX sources in archival {\it XMM-Newton} observations, 
but they did not succeed. We believe that it is
a matter of time until another pulsating ULX source is found
and IC10 X-1 is an intriguing candidate. 
But because such powerful X-ray emission ($L_X > 10^{40}~{\rm erg~s}^{-1}$)
from neutron stars must occur at
a favorable angle to the observer, we expect very few such pulsars 
to be discovered in the future \citep[see also][]{kin09}.

Two very recent studies of M82 X-2 have produced results 
that are different by an order of magnitude and on opposite sides 
of our result ($B_*\approx 3$~TG), but we understand the source of the differences.
In both studies, the mass accretion rate $\dot{M}$ at the disk's inner edge is too high.
In \cite{kin16}, the inner edge of the disk $R_d$ is located at $r_m$ which is 10 times smaller than corotation $r_c$. This causes highly supercritical accretion in the region between these two radii that can only occur if the magnetic field is weaker by a factor of 30. Thus, this model is not on the propeller line of M82 X-2.

In \cite{kar16}, the deduced $\dot{M}=3.2\times 10^{20}$~g~s$^{-1}$ is even more uncertain because it depends on a host of assumptions about the unknown parameters of the donor star and the spherically symmetric mass transfer between the two stars at large radii; it turns out to be $\sim$10 times larger than that determined in the \cite{kin16} model despite the fact that it was determined at corotation assuming that $r_c=r_m$. Thus, this model is not on the propeller line either.
In both cases, the Eddington value of $\dot{M}_{\rm Edd} = 2\times 10^{18}$~g~s$^{-1}$ applied at the disk's inner edge $R_d$ produces our result. Although it is not obvious how such a value can be deduced from the \cite{kar16} model (too many unknown parameters are involved), the \cite{kin16} model, which relies on the anisotropy of the X-ray emission, can be brought to agreement if the location of $R_d$ is moved closer to corotation.

\section*{Acknowledgments}
DMC and SGTL were supported by NASA grant NNX14-AF77G.
DK was supported by a NASA ADAP grant.

\label{lastpage}

\end{document}